\documentstyle[epsfig]{aipproc}
\newcommand{\be}{\begin{equation}}
\newcommand{\ee}{\end{equation}}
\newcommand{\bea}{\begin{eqnarray}}
\newcommand{\eea}{\end{eqnarray}}

\newcommand{\bm}[1]{\mbox{\boldmath $#1$}}

\newcommand{\xbj}{x_{\scriptscriptstyle B}}

\def\st{{\scriptscriptstyle T}}
\def\sL{{\scriptscriptstyle L}}
\def\slash{\rlap{/}}
\begin{document}
\title{Current fragmentation in\\[0.1cm] semiinclusive leptoproduction}

\author{P.J. Mulders}
\address{Division of Physics and Astronomy, Faculty of Exact
Sciences\\
Vrije Universiteit, De Boelelaan 1081\\
1081 HV Amsterdam, the Netherlands}

\maketitle

\begin{abstract}
Current fragmentation in semiinclusive deep inelastic leptoproduction
offers, besides refinement of inclusive measurements such as flavor
separation and access to the chiral-odd quark distribution functions 
$h_1^q(x) = \delta q(x)$, the possibility to investigate intrinsic transverse 
momentum of hadrons via azimuthal asymmetries.
\end{abstract}

\section*{Leading quark distribution functions}

In deep-inelastic leptoproduction (DIS), the {\em soft} hadron structure enters
via the quark distribution functions.
These distribution functions for a quark can be
obtained from the lightcone\footnote{
For inclusive leptoproduction the lightlike directions $n_\pm$ and lightcone
coordinates $a^\pm = a\cdot n_\mp$ are defined through hadron momentum $P$
and the momentum transfer $q$,
\begin{eqnarray*}
P = \frac{Q}{\xbj\sqrt{2}}\,n_+ + \frac{\xbj M^2}{Q\sqrt{2}}\,n_-,
\\
q = -\frac{Q}{\xbj\sqrt{2}}\,n_+ + \frac{Q}{\sqrt{2}}\,n_-.
\end{eqnarray*}
}
correlation functions~\cite{Soper77,Jaffe83,Manohar90,JJ92}.  
\be
\Phi_{ij}(x) = \left. \int \frac{d\xi^-}{2\pi}\ e^{ip\cdot \xi}
\,\langle P,S\vert \overline \psi_j(0) \psi_i(\xi)
\vert P,S\rangle \right|_{\xi^+ = \xi_\st = 0},
\label{eq1}
\ee
depending on the lightcone fraction of a quark (with momentum $p$),
$x = p^+/P^+$. In particular the 
At leading order, the relevant part of the correlator is $\Phi\gamma^+$ 
\bea
(\Phi\gamma^+)_{ij}
& = & \left. \int \frac{d\xi^-}{2\pi\sqrt{2}}\ e^{ip\cdot \xi}
\,\langle P,s^\prime\vert \psi^\dagger_{+j}(0) \psi_{+i}(\xi)
\vert P,s\rangle \right|_{\xi^+ = \xi_\st = 0}
\label{dens}
\eea
where $\psi_+ \equiv P_+\psi = \frac{1}{2}\gamma^-\gamma^+\psi$ is the
good component of the quark field~\cite{KS70}.

Explicitly, the matrix $M=(\Phi\gamma^+)^T$ in Dirac space
using a chiral representation becomes for a spin 0 target
the following 4 $\times$ 4 matrix,
\be
M_{ij} =
\left\lgroup \begin{array}{cccc}
f_1(x) & 0 & 0 & 0 \\
0 & 0 & 0 & 0 \\
0 & 0 & 0 & 0 \\
0 & 0 & 0 & f_1(x)
\end{array}\right\rgroup
\ee
In hard processes only two Dirac components are relevant, {\em one}
of them righthanded and {\em one} lefthanded ($\psi_{R/L} =
\frac{1}{2}(1\pm \gamma_5)\psi$). Restricting
ourselves to those states, the matrix for a spin 0 target
becomes 
\bea
M_{ij}
& = & \left\lgroup
\begin{array}{cc} f_1(x) & 0 \\
&\\
0 & f_1(x)
\end{array}\right\rgroup
\begin{array}{c}
\includegraphics[width = 0.45 cm]{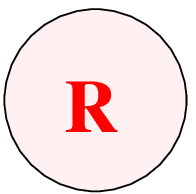}
\\[0.3cm] 
\includegraphics[width = 0.45 cm]{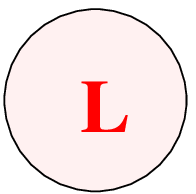}
\end{array}
\label{prod0}
\\
&&\mbox{}\hspace{0.8cm}
\includegraphics[width = 0.45 cm]{mulders-plenaryfig/helur.eps}
\hspace{0.9cm}\includegraphics[width = 0.45 cm]{mulders-plenaryfig/helul.eps}
\nonumber
\eea
For a spin 1/2 target more quark distributions appear in the lightcone
correlation function at leading order. In order to include all possible
target polarizations, one can employ a spin vector\footnote{
The spin vector is parametrized $S = S_\sL\,\frac{Q}{M\sqrt{2}}n_+
- S_\sL\,\frac{M}{Q\sqrt{2}}\,n_- + S_\st$.}, 
in which case one obtains \bea
M_{ij}
& = & \left\lgroup
\begin{array}{cc} f_1(x) + S_\sL\,g_1(x) & 0S_\st^1+i\,S_\st^2)\,h_1(x) \\
&\\
(S_\st^1-i\,S_\st^2)\,h_1(x) & f_1(x) + S_\sL\,g_1(x)
\end{array}\right\rgroup
\begin{array}{c}
\includegraphics[width = 0.45 cm]{mulders-plenaryfig/helur.eps}
\\[0.3cm] 
\includegraphics[width = 0.45 cm]{mulders-plenaryfig/helul.eps}
\end{array}
\\
&&\mbox{}\hspace{2.0cm}
\includegraphics[width = 0.45 cm]{mulders-plenaryfig/helur.eps}
\hspace{2.6cm}\includegraphics[width = 0.45 cm]{mulders-plenaryfig/helul.eps}
\nonumber
\eea
Equivalently, and for our purposes more instructive, one can also express $M$
as a $4 \times 4$ matrix in quark $\otimes$ nucleon spin space,
\bea
M^{\rm (prod)}\ & = &
\left\lgroup \begin{array}{cccc}
f_1 + g_1 & 0 & 0 & 2\,h_1 \\
& & &\\
0 & f_1 - g_1 & 0 & 0 \\
& & &\\
0 & 0 & f_1 - g_1 & 0 \\
& & &\\
2\,h_1 & 0 & 0 & f_1 + g_1
\end{array}\right\rgroup 
\ \begin{array}{c}
\includegraphics[width = 0.8 cm]{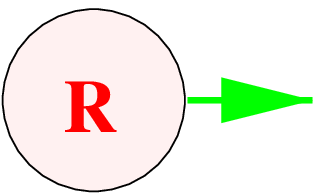}\\[0.3cm]
\includegraphics[width = 0.8 cm]{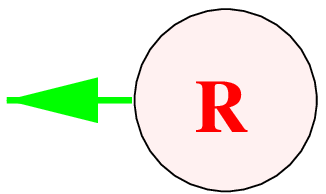}\\[0.3cm]
\includegraphics[width = 0.8 cm]{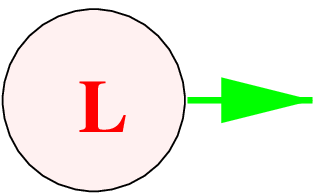}\\[0.3cm]
\includegraphics[width = 0.8 cm]{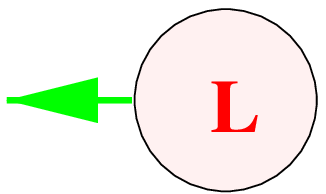}
\end{array}
\\
&&\mbox{}\hspace{0.7cm}
\includegraphics[width = 0.8 cm]{mulders-plenaryfig/helrr.eps}
\hspace{0.7cm}\includegraphics[width = 0.8 cm]{mulders-plenaryfig/hellr.eps}
\hspace{1.0cm}\includegraphics[width = 0.8 cm]{mulders-plenaryfig/helrl.eps}
\hspace{0.7cm}\includegraphics[width = 0.8 cm]{mulders-plenaryfig/helll.eps}
\nonumber
\eea
Note that the distribution functions exist for each quark flavor. The
functions are also denoted $f_1^q(x) = q(x)$, $g_1^q(x) = \Delta q(x)$
and $h_1^q(x) = \delta q(x)$. The three functions are independent.
From the fact that any forward matrix element of the above matrix represents a
density, one derives positivity bounds~\cite{Soffer95},
\bea
&& f_1(x) \ge 0 \\
&& \vert g_1(x)\vert \le f_1(x) \\
&& \vert h_1(x)\vert \le 
\frac{1}{2}\left( f_1(x) + g_1(x)\right) \le f_1(x).
\eea
As can be seen $h_1(x)$ involves a matrix elements between left- and
right-handed quarks, it is chirally odd~\cite{JJ92}. This implies that it is
not accessible in inclusive DIS, where the hard scattering part does not
change chirality except via (irrelevant) quark mass terms. 

By choosing a different basis of quark states, $\psi_{\uparrow/\downarrow}
= \frac{1}{2}\left( 1 + \gamma_5\gamma^1\right)\psi$ and nucleon transverse
spin states (along the x-axis),
\be
\vert N,\uparrow/\downarrow\rangle = 
\frac{1}{\sqrt{2}}\left( \vert N,+\rangle \pm \vert N,-\rangle \right),
\ee
one obtains the (equivalent) matrix
\bea
M^{\rm (prod)}\ & = &
\left\lgroup \begin{array}{cccc}
f_1 + h_1 & 0  & 0 & g_1 + h_1 \\
& & & \\
0 & f_1 - h_1 & g_1 - h_1 & 0 \\
& & & \\
0 & g_1 - h_1 & f_1 - h_1 & 0 \\
& & & \\
g_1 + h_1 & 0  & 0 & f_1 + h_1
\end{array}\right\rgroup 
\ \begin{array}{c}
\includegraphics[width = 0.45 cm]{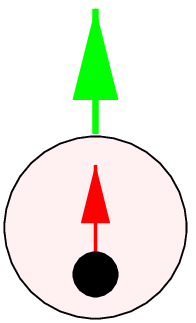}\\[0.1cm]
\includegraphics[width = 0.45 cm]{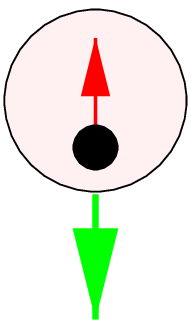}\\
\includegraphics[width = 0.45 cm]{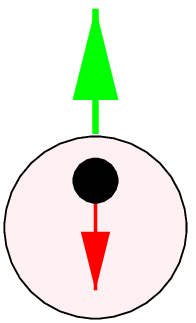}\\[0.1cm]
\includegraphics[width = 0.45 cm]{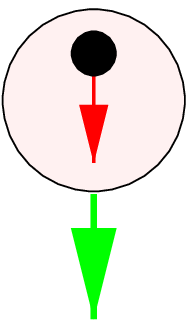}
\end{array}
\label{prod1}
\\
&&\mbox{}\hspace{0.8cm}
\includegraphics[width = 0.45 cm]{mulders-plenaryfig/helpp.eps}
\hspace{1.0cm}\includegraphics[width = 0.45 cm]{mulders-plenaryfig/helmp.eps}
\hspace{1.1cm}\includegraphics[width = 0.45 cm]{mulders-plenaryfig/helpm.eps}
\hspace{1.0cm}\includegraphics[width = 0.45 cm]{mulders-plenaryfig/helmm.eps}
\nonumber
\eea
from which one sees that $h_1(x)$ is a transverse spin density.

Leading gluon distribution functions correspond to lightcone correlators with
transverse gluon fields,
\be
\Gamma^{+\alpha;+\beta}(x)
= \left. \int \frac{d\xi^-}{2\pi}\ e^{ip\cdot \xi}
\,\langle P,S\vert F^{+\alpha}(0)F^{+\beta}(\xi)\vert P,S\rangle
\right|_{\xi^+ = \xi_\st = 0}.
\ee
This can be considered as a gluon production matrix, that for a spin 1/2
hadrons is given by 
\bea
M^{\rm (prod)}\ & = &
\left\lgroup \begin{array}{cccc}
G + \Delta G & 0 & 0 & 0 \\
& & &\\
0 & G - \Delta G & 0 & 0 \\
& & &\\
0 & 0 & G - \Delta G & 0 \\
& & &\\
0 & 0 & 0 & G + \Delta G
\end{array}\right\rgroup 
\ \begin{array}{c}
\includegraphics[width = 0.8 cm]{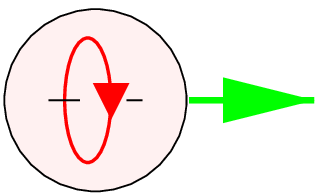}\\[0.3cm]
\includegraphics[width = 0.8 cm]{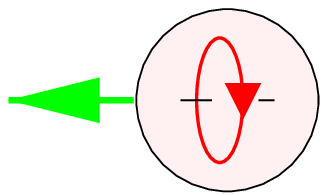}\\[0.3cm]
\includegraphics[width = 0.8 cm]{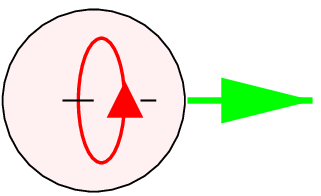}\\[0.3cm]
\includegraphics[width = 0.8 cm]{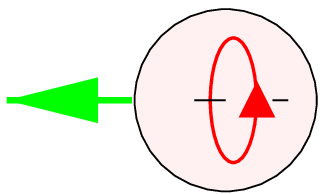}
\end{array}
\\
&&\mbox{}\hspace{0.9cm}
\includegraphics[width = 0.8 cm]{mulders-plenaryfig/helgpp.eps}
\hspace{0.8cm}\includegraphics[width = 0.8 cm]{mulders-plenaryfig/helgmp.eps}
\hspace{1.1cm}\includegraphics[width = 0.8 cm]{mulders-plenaryfig/helgpm.eps}
\hspace{0.8cm}\includegraphics[width = 0.8 cm]{mulders-plenaryfig/helgmm.eps}
\nonumber
\eea
Here we have used circularly polarized gluon states.

Inclusive DIS experiments have yielded a good knowledge of the unpolarized
quark distributions $q(x)$ in a nucleon and via the evolution equations of
$G(x)$. Polarized experiments have provided us with measurements of $\Delta
q(x)$ and first indications of $\Delta G(x)$. 

\section*{Semiinclusive leptoproduction}

Semiinclusive DIS (SIDIS), in particular one-particle inclusive DIS\footnote{
For SIDIS the lightlike directions $n_\pm$ and lightcone
coordinates $a^\pm = a\cdot n_\mp$ are defined through hadron momentum $P$
and $P_h$, in which case the momentum transfer $q$ requires a transverse
component 
\begin{eqnarray*}
P = \frac{Q}{\xbj\sqrt{2}}\,n_+ + \frac{\xbj M^2}{Q\sqrt{2}}\,n_-,
\\
q = -\frac{Q}{\sqrt{2}}\,n_+ + \frac{Q}{\sqrt{2}}\,n_- + q_\st,
\\
P_h = \frac{M_h^2}{Z_h\,Q\sqrt{2}}\,n_+ + \frac{z_h\,Q}{\sqrt{2}}\,n_-.
\end{eqnarray*}
},
can be and also has been used for additional flavor
identification. Instead of weighing quark flavors with the quark charge
squared $e_q^2$ one obtains a weighting with $e_q^2\,D_1^{q\rightarrow
h}(z_h)$, where $D_1^{q\rightarrow h}$ is the usual fragmentation function
for a quark of flavor $q$ into hadron $h$, experimentally accessible at $z_h$ =
$P\cdot P_h/P\cdot q$. 
possibilities to study intrinsic transverse momentum of partons, quarks and
gluons, via azimuthal asymmetries and the appearance of single spin
asymmetries via T-odd fragmentation functions.

\begin{figure} 
\centerline{
\epsfig{file=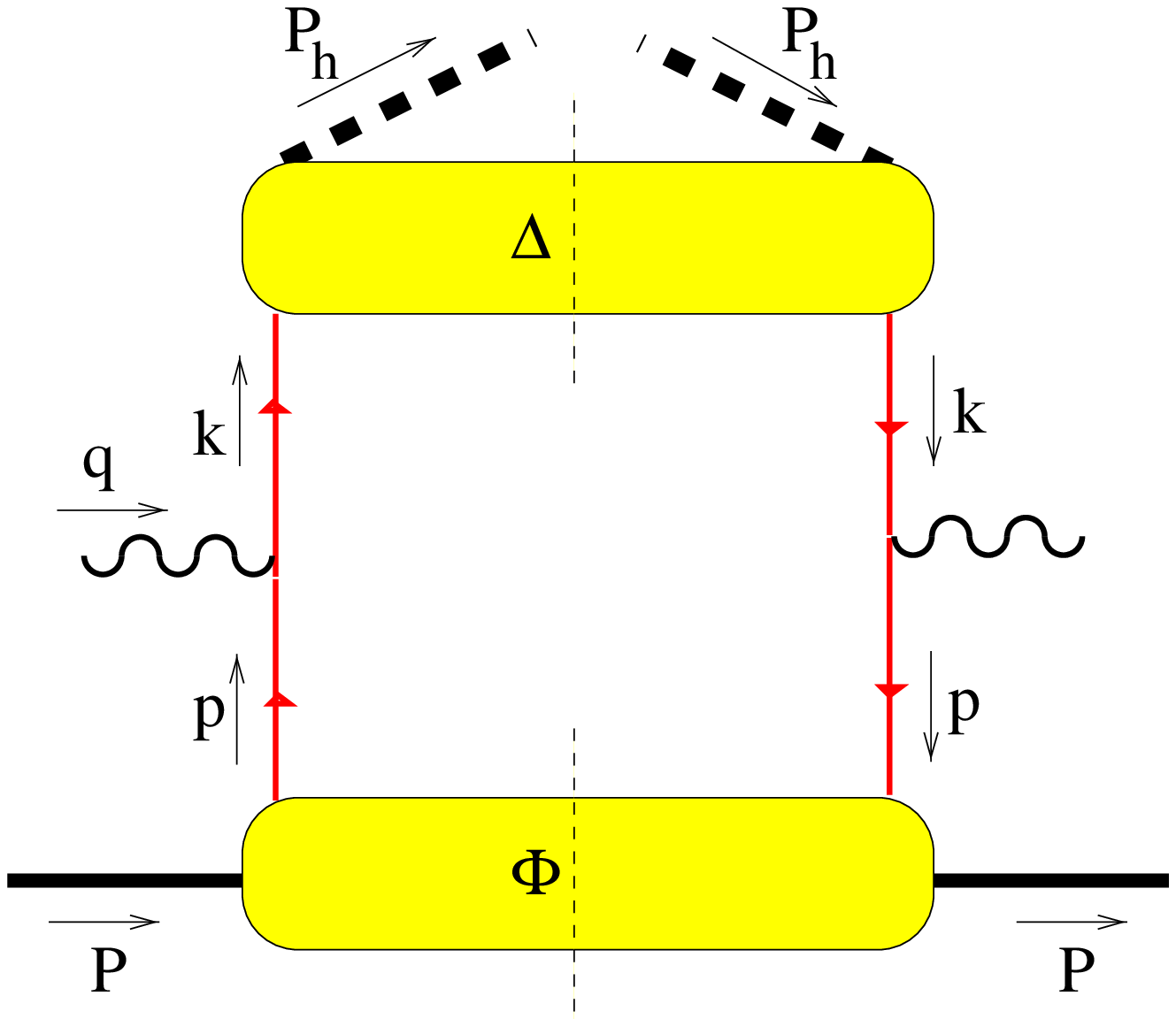,width=6cm} \hspace{1cm}
\epsfig{file=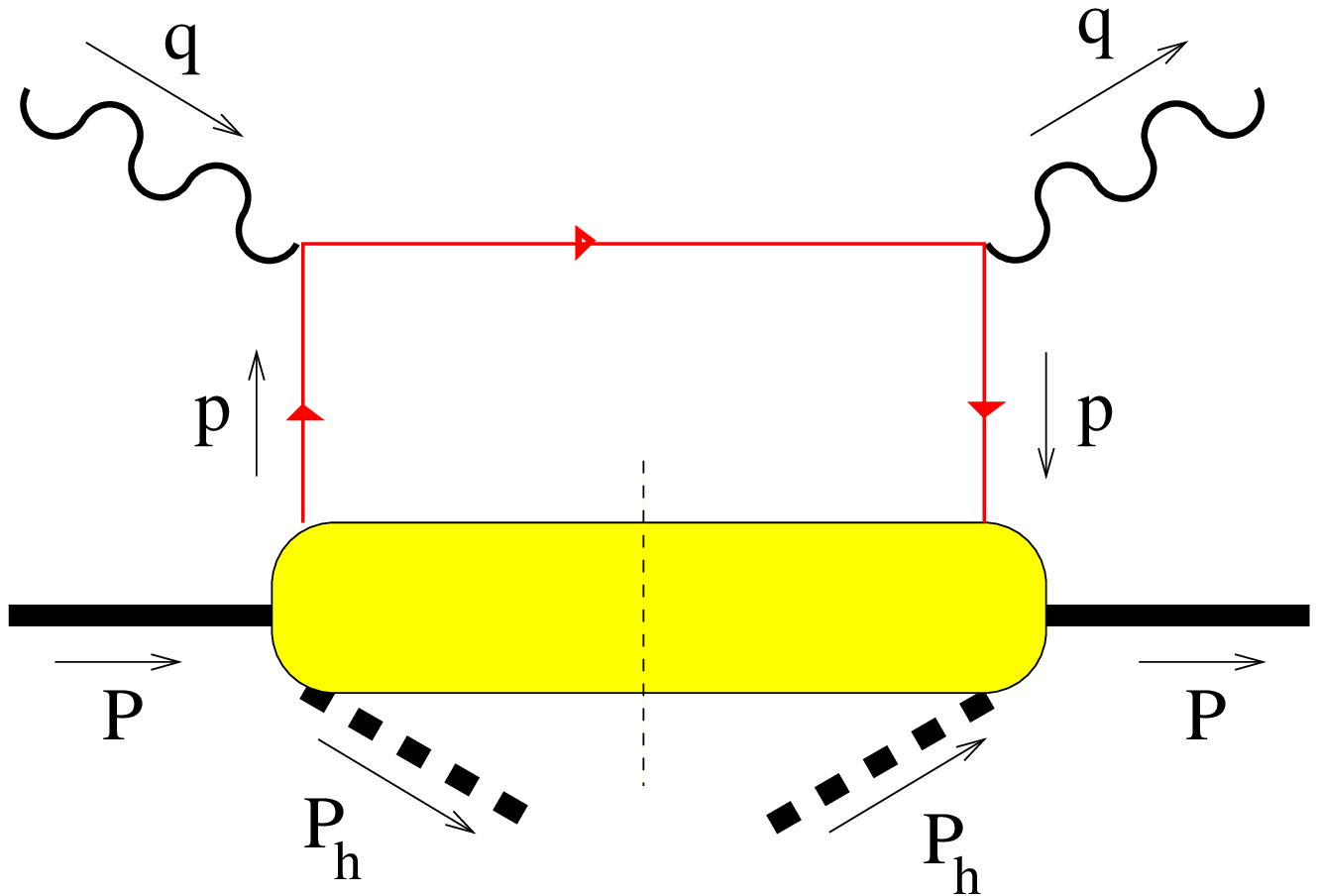, width=6cm}}
\vspace{10pt}
\caption{The leading contributions to current (left) and target (right)
fragmentation. \label{fig1}}
\end{figure}

Before turning to these topics, I want to address the issue of separation of
current fragmentation from target fragmentation, for which the leading order
description is illustrated in Fig.~\ref{fig1}.
While for current fragmentation we can use a description factorizing into
distribution and fragmentation functions, target fragmentation involves a
more complex soft part, namely fracture functions~\cite{fracture}.
Here we want to mention at least one check on the precision of current
fragmentation. Up to mass corrections of order $M^2/Q^2$ one has for current
fragmentation the identities 
\bea
&&
x = - \frac{q^+}{P^+} 
\approx \frac{Q^2}{2P\cdot q} 
\approx - \frac{P_h\cdot q}{P_h\cdot P},
\\
&&
z = \frac{P_h^-}{q^-} 
\approx  -\frac{2P_h\cdot q}{Q^2}
\approx \frac{P\cdot P_h}{P\cdot q}.
\eea
Actually incorporation of kinematical $1/Q^2$ corrections can be done by
calculating the lightcone ratios (first entries in both equations) in a frame
in which neither of the hadrons has a transverse momentum component.
\begin{figure}[b!]
\centerline{\epsfig{file=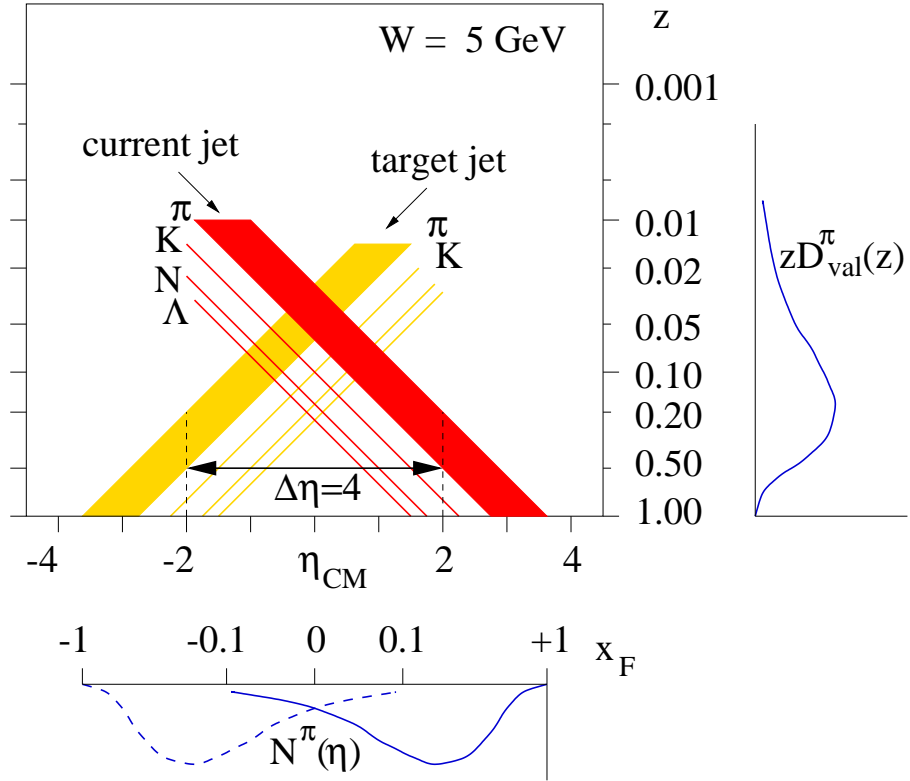,width=12cm}}
\vspace{10pt}
\caption{Relation between $z-values$ in fragmentation and CM rapidity for $W$ =
5 GeV.} \label{fig2}
\end{figure}
\begin{figure}[b!]
\centerline{\epsfig{file=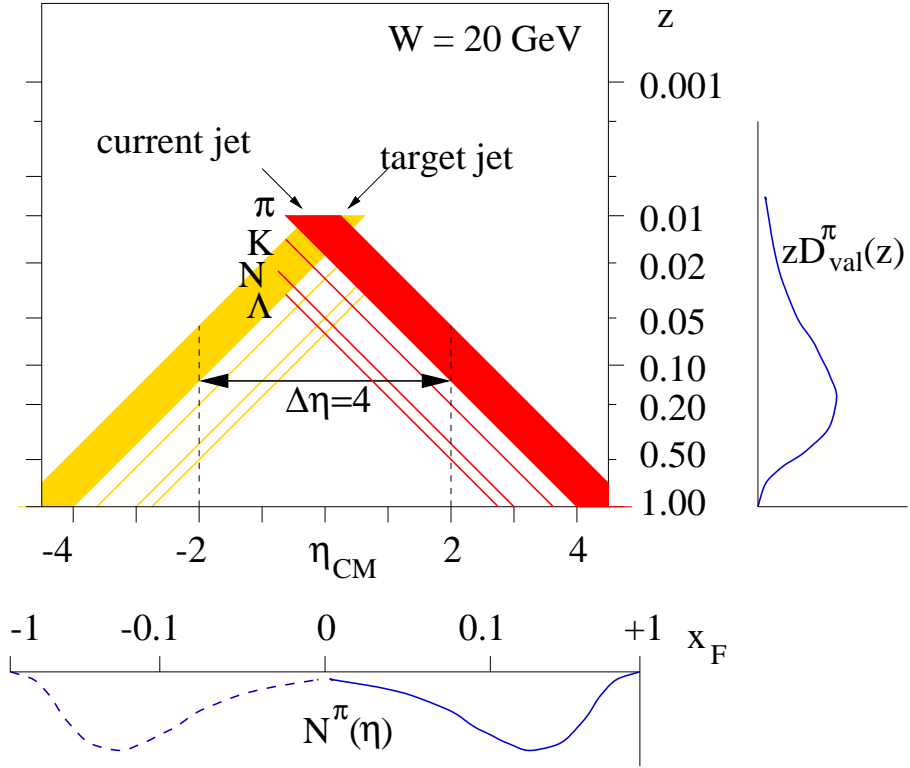,width=12cm}}
\vspace{10pt}
\caption{Relation between $z-values$ in fragmentation and CM rapidity for $W$
= 20 GeV.} \label{fig3}
\end{figure}

Based on results in the EMC compilation in ref.~\cite{Sloan} we take a
rapidity interval $\Delta \eta \approx 2$ (sometimes referred to as Berger's
criterium) to estimate the z-values for which one is most probably dealing
with current fragmentation. For this we construct a plot using the definition
of rapidity  
\be
\eta = \frac{1}{2}\,\ln\left(\frac{P_h^-}{P_h^+}\right) = 
\ln\left(\frac{P_h^-\sqrt{2}}{M_{h\perp}}\right) = 
-\ln\left(\frac{P_h^+}{M_{h\perp}}\right), 
\ee
where $M_{h\perp}^2 = M_h^2 + P_{h\perp}^2$. For current fragmentation one has
$z_c = P_h^-/q^-$ while for target fragmentation one is dealing with a
ratio $z_t = P_h^+/(1-x)P^+$. The proportionality is all we need to deduce
that for the center of mass rapidity one has
\bea
&&\eta_{cm} = \ln z_c + \ln\left(\frac{W}{M_{h\perp}}\right),
\\
&&\eta_{cm} = \ln z_t + \ln\left(\frac{W}{M_{h\perp}}\right),
\eea
where $W$ is the $\gamma^\ast N$ invariant mass, $W = (1-x)y\,s$, fixing the
maximum rapidity. For two values of $W$ = 5 and 20 GeV, we have indicated the
relation between $z$ and $\eta$ for both current and target fragmentation for
a number of hadrons in Figs~\ref{fig2} and \ref{fig3}. For light hadrons
the band reflects the influence of the transverse momentum. Looking at the
$\Delta \eta$ = 4 difference one can estimate $z$-values above which current
fragmentation dominates. Also indicated is how a typical (valence-like)
fragmentation function produces a number density in rapidity. Clearly seen is
how increased $W$ vastly lowers the $z$-values where one may expect to deal
with current fragments.

\section*{Leading quark distribition and fragmentation functions in SIDIS}

While the distribution functions in DIS could be obtained from the lightcone
correlation function in Eq.~\ref{eq1}, one encounters in SIDIS two types of
lightfront correlation functions, involving also transverse momenta of
partons as first pointed out by Ralston and Soper~\cite{RS79,TM95}
One part is relevant to treat quarks in a hadron 
\be 
\Phi_{ij}(x,\bm p_T) =
\left. \int \frac{d\xi^-d^2\bm \xi_T}{(2\pi)^3}\ e^{ip\cdot \xi}
\,\langle P,S\vert \overline \psi_j(0) \psi_i(\xi)
\vert P,S\rangle \right|_{\xi^+ = 0},
\label{phi}
\ee
depending on $x=p^+/P^+$ and the quark transverse 
momentum $\bm p_\st$ in a target with $P_\st = 0$.
A second correlation function~\cite{CS82}
\be
\Delta_{ij}(z,\bm k_\st) =
\left. \sum_X \int \frac{d\xi^-d^2\bm \xi_\st}{(2\pi)^3} \,
e^{ik\cdot \xi} \langle 0 \vert \psi_i (\xi) \vert P_h,X\rangle
\langle P_h,X\vert\overline \psi_j(0) \vert 0 \rangle
\right|_{\xi^+ = 0},
\label{delta}
\ee
describes fragmentation of a quark into a hadron. It depends
on $z = P_h^+/k^+$ and the quark transverse momentum $k_\st$ when one
produces a hadron with $P_{h\st} = 0$. A simple boost shows  that this is
equivalent to a quark producing a hadron with transverse  momentum $P_{h\perp}
= -z\,k_\st$ with respect to the quark. 

As before we make the Dirac structure explicit and find at leading order only
two relevant components, one of them righthand and one lefthanded. For
fragmentation into spin 0 hadrons (e.g. pion production) this leads to
the following 2 $\times$ 2 quark decay matrix,
\bea
M^{\rm (decay)}_{ij}
& = & \left\lgroup
\begin{array}{cc} D_1(z) & 
i\frac{\vert k_\st\vert\,e^{i\phi}}{M_h}\,H_1^\perp(z) \\ &\\
-i\frac{\vert k_\st\vert\,e^{i\phi}}{M_h}\,H_1^\perp(z) & D_1(z)
\end{array}\right\rgroup
\begin{array}{c}
\includegraphics[width = 0.45 cm]{mulders-plenaryfig/helur.eps}
\\[0.5cm] 
\includegraphics[width = 0.45 cm]{mulders-plenaryfig/helul.eps}
\end{array}
\\
&&\mbox{}\hspace{1.8cm}
\includegraphics[width = 0.45 cm]{mulders-plenaryfig/helur.eps}
\hspace{2.5cm}\includegraphics[width = 0.45 cm]{mulders-plenaryfig/helul.eps}
\nonumber
\eea
As compared to the production matrix in Eq.~\ref{prod0} one has the additional
function $H_1^\perp(z)$, which is allowed because one cannot use time-reversal
invariance to constrain the structure of $\Delta$ in Eq.~\ref{delta}. Such
functions are referred to as T-odd. We note, furthermore, that $H_1^\perp$ is
also chiral-odd, hence it will appear in a cross section in combination with a
chiral-odd distribution function such as $h_1$. The appearance of the 
transverse momentum, however, has as consequence that this fragmentation
function only can be measured via the dependence on the 
transverse momentum of the produced hadron, e.g. in azimuthal asymmetries.

For a spin 1/2 hadron one finds that the structure of $\Phi$ including
transverse momentum dependence, leads to the production matrix,
\be
M^{\rm (prod)}=
\left\lgroup \begin{array}{cccc}
f_1 + g_{1} &
\frac{\vert p_\st\vert}{M}\,e^{i\phi}\,g_{1T}&
\frac{\vert p_\st\vert}{M}\,e^{-i\phi}\,h_{1L}^\perp&
2\,h_{1} \\
& & & \\
\frac{\vert p_\st\vert}{M}\,e^{i\phi}\,g_{1T}^\ast&
f_1 - g_{1} &
\frac{\vert p_\st\vert^2}{M^2}e^{-2i\phi}\,h_{1T}^\perp &
-\frac{\vert
p_\st\vert}{M}\,e^{-i\phi}\,h_{1L}^{\perp\ast}\\ & & &
\\ \frac{\vert p_\st\vert}{M}\,e^{i\phi}\,h_{1L}^{\perp\ast}&
\frac{\vert p_\st\vert^2}{M^2}e^{2i\phi}\,h_{1T}^\perp &
f_1 - g_{1} &
-\frac{\vert p_\st\vert}{M}\,e^{i\phi}\,g_{1T}^\ast \\
& & & \\
 2\,h_{1}&
-\frac{\vert p_\st\vert}{M}\,e^{i\phi}\,h_{1L}^\perp&
-\frac{\vert p_\st\vert}{M}\,e^{-i\phi}\,g_{1T} &
f_1 + g_{1}
\end{array}\right\rgroup ,
\ee
to be compared with Eq.~\ref{prod1}. Using time-reversal invariance all the
distribution functions appearing in this equation are expected to be real,
leaving aside mechanisms discussed in Refs~\cite{Sivers90}. For fragmentation
functions, however, T-reversal cannot be used~\cite{RKR71,HHK83,JJ93}, leading
to two T-odd fragmentation functions~\cite{Collins93,MT96}. They are the
imaginary parts of the complex off-diagonal ($p_\st$-dependent) functions. To
be precise one obtains the decay matrix with fragmentation functions after the
replacements $f_1 \rightarrow D_1$, $g_1 \rightarrow G_1$, $h_1 \rightarrow
H_1$, $g_{1T} \rightarrow G_{1T} + i\,D_{1T}^\perp$ and  $h_{1L}^\perp
\rightarrow H_{1L}^\perp + i\,H_1^\perp$.
\\[0.3cm]
{\em The possibility to access the full (transverse momentum dependent) spin
structure of the nucleon is in my opinion one of the most exciting
possibilities offered by 1-particle inclusive leptoproduction.}

\subsection*{Bounds}
In analogy to the Soffer bound derived from the production matrix in
Eq.~\ref{prod0} one easily derives a number of new bounds from the full
matrix, such as
\bea
&&f_1(x,\bm p_\st^2) \ge 0 \, , 
\\
&&\vert g_{1}(x,\bm p_\st^2)\vert \le f_1(x,\bm p_\st^2) \,.
\eea
obtained from one-dimensional subspaces
and
\bea
&& \vert h_1 \vert \le
\frac{1}{2}\left( f_1 + g_{1}\right)
\le f_1,
\label{Soffer}\\
&&
\vert h_{1T}^{\perp(1)}\vert \le
\frac{1}{2}\left( f_1 - g_{1}\right)
\le f_1,
\\
&& \vert g_{1T}^{(1)}\vert^2
\le \frac{\bm p_\st^2}{4M^2}
\left( f_1 + g_{1}\right)
\left( f_1 - g_{1}\right)
\le \frac{\bm p_\st^2}{4M^2}\,f_1^2,
\\
&& \vert h_{1L}^{\perp (1)}\vert^2
\le \frac{\bm p_\st^2}{4M^2}
\left( f_1 + g_{1}\right)
\left( f_1 - g_{1}\right)
\le \frac{\bm p_\st^2}{4M^2}\,f_1^2,
\eea
obtained from two-dimensional subspaces.
Here we have introduced the notation $g_{1T}^{(1)}(x,\bm p_\st^2) \equiv
(\bm p_\st^2/4M^2)\,g_{1T}(x,\bm p_\st^2)$.
These bounds and their further refinements have been discussed in detail
in Ref.~\cite{BBHM}. There are straightforward extensions of transverse
momentum dependent distribution and fragmentation functions for spin 1 
hadrons~\cite{bacchetta} and gluons in spin 1/2 hadrons~\cite{rodrigues}.

\subsection*{Bound on the Collins function}

As an application of using the bounds, consider the Collins function
$H_1^{\perp (1)}$ for which we have
\be
\vert H_1^{\perp (1)}(z,-z\bm k_\st)\vert 
= \vert \frac{\bm k_\st^2}{2M_\pi^2}\,H_1^{\perp}(z,-z\bm k_\st)\vert
\le \frac{\vert k_\st\vert}{2M_\pi}\,D_1(z,-z\bm k_\st) .
\ee
With the assumption
\be
D_1(z,-z\bm k_\st) 
= D_1(z)\,\frac{R_\pi^2(z)}{\pi\,z^2}\,e^{-\vert k_\st\vert^2\,R_\pi^2},
\ee
one finds for the function integrated over transverse momenta,
\be
\vert H_1^{\perp (1)}(z)\vert \le
\underbrace{\frac{\sqrt{\pi}}{4M_\pi R_\pi(z)}}_{{\cal O}(1)}\,D_1(z) .
\ee

\subsection*{Lorentz invariance relations}
Since both the $p_\st$ integrated functions and the $p_\st$ dependent functions
originate from (nonlocal) combinations of two quark fields, Poincar\'e
invariance poses restrictions on the various ways we project out distribution
functions. In particular we consider the inclusion of the higher-twist
functions for the $p_\st$-integrated functions, in which case the correlator
$\Phi(x)$ in Eq.~\ref{eq1} becomes~\cite{JJ92} 
\bea
\Phi(x) & = &
\frac{1}{2}\Biggl\{ 
f_1\slash n_+ 
+ S_\sL\,g_{1}\gamma_5\slash n_+
+ h_{1}\,\frac{[\slash S_\st,\slash n_+]\gamma_5}{2}
\Biggr\} \nonumber
\\ && \mbox{}+\frac{M}{2P^+}\Biggl\{
e + g_T\,\gamma_5\slash S_\st
+ S_\sL\,h_L\,\frac{[\slash n_+,\slash n_-]\gamma_5}{2}
\Biggr\} \nonumber
\\ && \mbox{}+\frac{M}{2P^+}\Biggl\{
f_T\,\epsilon_\st^{\rho\sigma}S_{\st\rho}\gamma_\sigma
-i\,S_\sL\,e_L\,\gamma_5 
+ \,h\,\frac{i\,[\slash n_+,\slash n_-]}{2}
\Biggr\} .
\label{phi2}
\eea
\mbox{}\\[-1.4cm]
\mbox{}\hspace{1.5cm}
\includegraphics[width=0.8 cm]{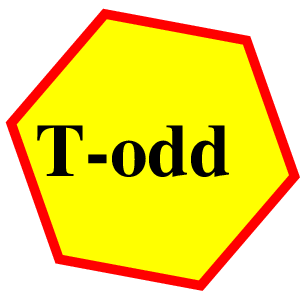}
\\[0.3cm]
We will compare this with the $p_\st$-integrated
result after weighing the $\Phi(x,p_\st)$ with $p_\st$,
giving $\Phi_\partial^\alpha (x) 
\equiv \int d^2p_\st\ p_\st^\alpha\,\Phi(x,p_\st)$, explicitly
\bea
\frac{1}{M}\,\Phi_\partial^{\alpha}(x) & = &
\frac{1}{2}\,\Biggl\{
g_{1T}^{(1)}\,S_\st^\alpha\,\gamma_5\slash n_+
+S_\sL\,h_{1L}^{\perp (1)}
\,\frac{[\slash n_+, \gamma^\alpha]\gamma_5}{2}
\nonumber \\ && \mbox{}
\quad +f_{1T}^{\perp(1)}\,\epsilon_\st^{\alpha\beta}
S_{\st\beta} \slash n_+  
+h_1^{\perp (1)}\,\frac{i\,[\slash n_+, \gamma^\alpha]}{2}
\Biggr\} .
\eea
\mbox{}\\[-1.4cm]\mbox{}\hspace{1.5cm}
\includegraphics[width=0.8 cm]{mulders-plenaryfig/Toddmarker.eps}
\\[0.3cm]
The $\bm p_\st^2/2M^2$ moment of the transverse momentum dependent functions
turn out to be related to twist-three functions~\cite{BKL,BM98,MT96},
\bea
&&\underbrace{g_T - g_1}_{g_2} = \frac{d}{dx}\,g_{1T}^{(1)},
\\ &&
\underbrace{h_L - h_1}_{{1\over 2}\,h_2}
= -\,\frac{d}{dx}\,h_{1L}^{\perp(1)},
\\ && 
f_T = -\,\frac{d}{dx}\,f_{1T}^{\perp(1)},
\\ &&
h = -\,\frac{d}{dx}\,h_1^{\perp(1)}.
\eea
The above relations can for instance be used to estimate the magnitude of
$g_{1T}^{(1)}$ from polarized inclusive data on $g_2$~\cite{KM,BM00}.

\section*{Azimuthal asymmetries}

As already mentioned before, in order to experimentally investigate the full
spin structure including the off-diagonal transverse momentum dependent
functions (Eq.~\ref{prod1}) one needs semiinclusive measurements. The
transverse momentum dependence is probed via specific azimuthal asymmetries.
We limit ourselves here to just one example, but before doing so remind the
reader of the 'rules'. 
\begin{itemize}
\item
Depending on the powers $t$ of $(M/P^+)$ [for
fragmentation functions powers of $(M_h/P_h^-)$] the functions show up in
contributions in the cross section behaving as $(1/Q)^t$. This is sometimes
referred to as a twist expansion, although it in particular for the transverse
momentum dependent correlators $\Phi(x,p_\st)$ and $\Delta(z,-zk_\st)$ only
indicates the 'lowest twist' operators that play a role, now using twist in
the rigorous operator-product-expansion sense.
\item
Cross sections are chirally even. For instance chirally even functions like
$f_{\ldots}$ or $g_{\ldots}$ appear together with chirally even fragmentation
functions $D_{\ldots}$ or $G_{\ldots}$, while chirally odd functions
$h_{\ldots}$ and $e$ appear together with chirally odd functions $H_{\ldots}$
and $E$. Note that terms originating from quark mass terms multiply
combinations of opposite chirality.
\item
The number of polarizations needed is even in the case of an even
number of $T-odd$ functions combinations of distribution and fragmentation
functions and it is odd in the case of an odd number of $T-odd$ functions.
\end{itemize}
The following explicit example serves to illustrate these points, namely the
semi-inclusive asymmetry
\be
\left<\frac{Q_\st}{M_\pi}\,\sin(\phi^\ell_h+\phi^\ell_S)\right>_{OTO}
= \frac{2\pi \alpha^2\,s}{Q^4}\,\vert \bm S_\st \vert
\,2(1-y)\sum_{a,\bar a} e_a^2
\,\xbj\,h_1^a(\xbj) H_1^{\perp(1)a}(z_h),
\ee
which is the cross section weighted with the magnitude $Q_T = \vert
P_{h\perp}/z_h$ and involving the angles of the transverse momentum of the
produced hadron, $\phi_h^\ell$ (with repect to lepton scattering plane) and
the transverse spin of the target, $\phi_S^\ell$. Since the Collins functions
$H_1^\perp$ is T-odd and chirally odd, it can appear together with the
chirally odd distribution function $h_1$, but since the latter is T-even, the
combination appears in a single spin asymmetry: unpolarized lepton,
transversely polarized target, production of a spinless particle.
Many other examples have been discussed in the
literature~\cite{KT,MT96,BM98,BJM}, some of them will be discussed at this
meeting~\cite{Bog}. Also recent experimental indications of nonvanishing
azimuthal asymmetries exist~\cite{SMC,HERMES,LEP}

\section*{QCD dynamics}

The study of distribution and fragmentation functions is interesting since it
identifies well-defined quantities that can be extracted from experiment by
using high energy (expansion in powers of $1/Q$ with calculable $\ln Q^2$
perturbative corrections) and identified as specific matrix elements of quark
and gluon fields. We will illustrate below how the QCD
dynamics enters here. In Eq.~\ref{phi2} the quark-quark correlation function
was expanded including (higher-twist) terms proportional to $(M/P^+)$. These
terms are the leading terms in a correlator involving $\overline
\psi(0)\,D^\alpha \psi(\xi)$ where $D^\alpha$ is the covariant derivative. For
transverse indices one can use the QCD equations of motion, $(i\slash D -
m)\psi = 0$ to show that
\bea
\frac{1}{M}\,\Phi_D^{\alpha}(x) & = &
\frac{1}{2}\,\Biggl\{
\left(xg_T -{m\over M}\,h_1\right)
\,S_\st^\alpha\,\gamma_5\slash n_+
+ S_\sL \left(x h_L  - {m\over M}\,g_1\right)
\,\frac{[\slash n_+, \gamma^\alpha]\gamma_5}{2}
\nonumber \\ && \mbox{}\quad - xf_T
\ \epsilon_\st^{\alpha\beta} S_{\st\beta}\slash n_+  
- xh\ \frac{i\,[\slash n_+, \gamma^\alpha]}{4} \Biggr\}
\eea
\mbox{}
\\[-1.4cm]
\mbox{}\hspace{1.5cm}
\includegraphics[width=0.8 cm]{mulders-plenaryfig/Toddmarker.eps}
\\[0.3cm]
One can identify the socalled {\em interaction dependent} pieces
via $\Phi_A \equiv \Phi_D - \Phi_\partial$
\bea
\frac{1}{M}\,\Phi_A^{\alpha}(x) & = &
\frac{1}{2}\,\Biggl\{
\underbrace{\left(xg_T - g_{1T}^{(1)}-{m\over M}\,h_1
\right)}_{x\tilde g_T}
\,S_\st^\alpha\,\gamma_5\slash n_+
\nonumber \\ && \mbox{}
\qquad + S_\sL \underbrace{\left(xh_L - h_{1L}^{\perp (1)} - {m\over M}\,g_1
\right)}_{x\tilde h_L}
\,\frac{[\slash n_+, \gamma^\alpha]\gamma_5}{2}
\nonumber \\ && \mbox{}
\qquad -\underbrace{\left(xf_T + f_{1T}^{\perp(1)}\right)}_{x\tilde f_T}
\,\epsilon_\st^{\alpha\beta}S_{\st\beta} \slash n_+  
-\underbrace{\left( xh + 2h_1^{\perp (1)}\right)}_{x\tilde h}
\,\frac{i\,[\slash n_+, \gamma^\alpha]}{4}\Biggr\}
\eea
\mbox{}
\\[-1.8cm]
\mbox{}\hspace{1.5cm}
\includegraphics[width=0.8 cm]{mulders-plenaryfig/Toddmarker.eps}

\subsection*{Relations}
The relations following from Lorentz invariance and the equations of motion
can be combined to relate the functions discussed above.
In particular consider the 'leading' functions $g_1$ and
$g_{1T}^{(1)}$ appearing in the matrix in Eq.~\ref{prod1} and the 
'subleading' functions $g_T$ and $\tilde g_T$ discussed in the previous
section. From  the equations of motion and Lorentz invariance, respectively, we
get (omitting quark mass terms),  
\bea
g_T & = & \frac{g_{1T}^{(1)}}{x} + \tilde g_T
\nonumber
\\
& = & g_1 + \frac{d}{dx}\,g_{1T}^{(1)}
\eea
from which it is straightforward to derive the Wandzura-Wilczek 
relation~\cite{WW,MT96},
\be
g_T = \int_x^1 dy\,\frac{g_1(y)}{y} 
+ \underbrace{\left(\tilde g_T - 
\int_x^1 dy\,\frac{\tilde g_T(y)}{y}\right)}_{\bar g_T}.
\ee
Using this also $g_{1T}^{(1)}$ can be expressed in $g_1$ and $\tilde g_T$.
Often this relation is used to make further assumptions, e.g. the assumption
that the interaction-dependent part $\tilde g_T \approx 0$ (and hence also
$\bar g_T \approx 0$) or the assumption that the $p_\st$-weighted function
$g_{1T}^{(1)} \approx 0$. Although such assumptions are at the present time
still fairly {\em ad hoc}, they allow us to obtain order of magnitude 
estimates of the functions from just the leading twist function $g_1$.

The equivalent relations for the $h$-functions are
\bea
h_L & = & -2\,\frac{h_{1L}^{\perp(1)}}{x} + \tilde h_L
\nonumber \\
& = & h_1 - \frac{d}{dx}\,h_{1L}^{\perp(1)},
\eea
from which one obtains~\cite{JJ92,MT96}
\be
h_L  = 2x\int_x^1 dy\,\frac{h_1(y)}{y^2} + \underbrace{\left(\tilde h_L - 
2x\int_x^1 dy\,\frac{\tilde h_L(y)}{y^2}\right)}_{\bar h_L}.
\ee

For the T-odd functions, we will present the equivalent relations for the
Collins fragmentation functions, 
\bea
H(z) & = & -2z\,H_{1}^{\perp(1)}(z) + \tilde H(z)
\nonumber \\
& = & z^3\,\frac{d}{dz}\left(\frac{H_1^{\perp(1)}}{z}\right),
\eea
from which one obtains~\cite{MT96}
\be
H(z) = \tilde H(z) + 2\int_z^1 dz^\prime \ \frac{\tilde
H(z^\prime)}{z^\prime} ,
\ee
i.e. this T-odd function is purely interaction-dependent as one might have
expected for such functions.

\section*{Concluding remarks}

In this talk I have tried to indicate new opportunities in semiinclusive
leptoproduction. In a collider with sufficient energy one can reliably study
current fragmentation. This allows first of all a better flavor separation of
the 'ordinary' unpolarized and polarized distribution functions $f_1^q(x)$ and
$g_1^q(x)$. In principle, it also allows access to the chiral-odd distribution
function $h_1^q(x)$, but the measurements require a chiral-odd
fragmentation function, which for the case that one integrates over all
transverse momenta, requires polarimetry in the final state. Measurement of
transverse momenta of the produced hadron opens a rich new field, e.g. the
existence of chiral-odd fragmentation function $H_1^\perp$ for spin 0
particles (Collins function). Since this function is also T-odd, it enables
access to $h_1^q$ via a single spin asymmetry. Last but not least one must
realize that the transverse momentum dependent functions carry the
information on the nonperturbative structure of the nucleon, often in a way
complimentary to higher-twist functions.


\begin{references}
\bibitem{Soper77}
D.E. Soper,
Phys. Rev.~D 15 (1977) 1141;
Phys. Rev. Lett.  43 (1979) 1847.
\bibitem{Jaffe83}
R.L. Jaffe,
Nucl. Phys.~B 229 (1983) 205.
\bibitem{Manohar90}
A.V. Manohar, Phys. Rev. Lett.~65 (1990) 2511.
\bibitem{JJ92}
R.L. Jaffe and X. Ji,
Nucl. Phys.~B 375 (1992) 527.
\bibitem{KS70}
J.B. Kogut and D.E. Soper, Phys. Rev.~D 1 (1970) 2901.
\bibitem{Soffer95}
J. Soffer, Phys. Rev. Lett.~74 (1995) 1292.
\bibitem{fracture}
L. Trentadue and G. Veneziano, Phys. Lett.~B 323 (1994) 201;
D. de Florian and R. Sassot, Phys. Rev.~D 56 (1997) 426;
M. Grazzini, G.M. Shore and B.E. White, Nucl. Phys.~B 555 (1999) 259.
\bibitem{Sloan}
T. Sloan, G. Smadja and R. Voss, Phys. Rep.~162 (1988) 45.
\bibitem{RS79}
J.P. Ralston and D.E. Soper, Nucl. Phys.~B 152 (1979) 109.
\bibitem{TM95}
R. D. Tangerman and P.J. Mulders, Phys. Rev.~D 51 (1995) 3357
\bibitem{CS82}
J.C. Collins and D.E. Soper, Nucl. Phys.~B 194 (1982) 445.
\bibitem{Sivers90}
Possible T-odd effects could arise from soft initial state interactions
as outlined in 
D. Sivers, Phys. Rev.~D 41 (1990) 83 and Phys. Rev.~D 43 (1991) 261
and M. Anselmino, M. Boglione and F. Murgia, Phys. Lett.~B 362 (1995) 164.
Also gluonic poles might lead to presence of T-odd functions, see
N. Hammon, O. Teryaev and A. Sch\"afer, Phys.~Lett.~B 390 (1997) 409
and
D. Boer, P.J. Mulders and O.V. Teryaev, Phys.~Rev.~D 57 (1998) 3057.
\bibitem{RKR71}
A. De R\'ujula, J.M. Kaplan and E. de Rafael, 
Nucl. Phys.~B 35 (1971) 365.
\bibitem{HHK83}
K. Hagiwara, K. Hikasa and N. Kai, Phys. Rev.~D 27 (1983) 84.
\bibitem{JJ93}
R.L. Jaffe and X. Ji, Phys. Rev. Lett.~71 (1993) 2547.
\bibitem{Collins93}
J. Collins, Nucl. Phys.~B 396 (1993) 161.
\bibitem{MT96}
P.J. Mulders and R.D. Tangerman, Nucl.~Phys.~B 461 (1996) 197;
Nucl.~Phys.~B 484 (1997) 538 (E).
\bibitem{BBHM}
A. Bacchetta, M. Boglione, A. Henneman and P.J. Mulders, Phys. Rev. Lett.~58
(2000) 712 
\bibitem{bacchetta}
A. Bacchetta and P.J. Mulders, hep-ph/0007120.
\bibitem{rodrigues}
P.J. Mulders and J. Rodrigues, hep-ph/0009343.
\bibitem{BKL}
A.P. Bukhvostov, E.A. Kuraev and L.N. Lipatov, Sov. Phys. JETP~60
(1984) 22.
\bibitem{BM98}
D. Boer and P.J. Mulders, Phys. Rev.~D 57 (1998) 5780.
\bibitem{KM}
A.M. Kotzinian and P.J. Mulders, Phys. Rev.~D 54 (1996) 1229; 
A.M. Kotzinian and P.J. Mulders, Phys. Lett.~B 406 (1997) 373.
\bibitem{BM00}
P.J. Mulders and M. Boglione, Nucl. Phys. A~666\&667 (2000) 257c.
\bibitem{KT}
A.M. Kotzian, Nucl. Phys.~B 441 (1995) 234;
R.D. Tangerman and P.J. Mulders, Phys. Lett.~B 352 (1995) 129.
\bibitem{BJM}
D. Boer, R. Jakob and P.J. Mulders, Nucl. Phys.~B 564 (2000) 471.
\bibitem{Bog}
M. Boglione, contribution to this meeting, hep-ph/0010166;
M. Boglione and P.J. Mulders, Phys. Rev.~D 60 (1999) 054007;
M. Boglione and P.J. Mulders, Phys. Lett.~B 478 (2000) 114.
\bibitem{SMC}
A. Bravar, Nucl. Phys. Proc. Suppl.~B 79 (1999) 521.
\bibitem{HERMES}
H. Avakian, Nucl. Phys. Proc. Suppl.~B 79 (1999) 523.
\bibitem{LEP}
E. Efremov, O.G. Smirnova and L.G. Tkatchev, 
Nucl. Phys. Proc. Suppl.~B 79 (1999) 554.
\bibitem{WW}
S. Wandzura and F. Wilczek, Phys. Rev.~D 16 (1977) 707.
\end{references}
\end{document}